\documentstyle[titlepage,12pt]{article}
\begin{document}
\begin{titlepage}
\begin{flushright}
PITHA 95/30\\
October 95
\end{flushright}
\vspace{0.8cm}
\begin{center}
{\bf\LARGE Probing Higgs Boson - and
Supersymmetry-Induced CP Violation }
{\bf\LARGE in Top Quark Production by (Un-) Polarized
Electron-Positron Collisions }

\vspace{2cm}
\centerline{ \bf Werner Bernreuther$^a$
\footnote{Research supported  by BMBF contract 056AC92PE} and Peter
Overmann$^b$}
\vspace{1cm}
\centerline{$^a$Institut f. Theoretische Physik,
RWTH, D-52056 Aachen, Germany}
\centerline{$^b$Institut f. Physik,
Universit\"at Dortmund, D-44221 Dortmund , Germany}
\vspace{3cm}

{\bf Abstract:}\\
\parbox[t]{\textwidth}
{We consider, both for unpolarized and longitudinally polarized
electron beams, the reaction  $\rm e^+ e^- \to  t\bar t$ with
subsequent semileptonic $t$ and nonleptonic $\bar t$ decay  and vice
versa and investigate optimized  angular correlations which are sensitive
to CP non-conservation in the $t \bar t$ production vertex.   We
calculate these correlations for  two-Higgs--doublet extensions and
the minimal supersymmetric extension of the Standard Model (SM)  with
CP violation beyond the Kobayashi-Maskawa phase. While the
sensitivity of the optimal correlation for tracing dispersive  CP
effects is enhanced  with longitudinally  polarized electron beams,
we find that the sensitivity of  the best correlation for probing
absorptive CP effects is almost independent of the polarization
degree.
}  \end{center}
\end{titlepage}
\newpage

\bibliographystyle{unsrt}
\bigskip
\section{Introduction}
\noindent One of the  pillars of the physics program of a future
high luminosity linear
electron positron collider would be the detailed investigation of top
quarks \cite{Zerwas}.
The production and decay of these quarks could be studied at such a facility
under rather clean
conditions. There are quite a number of proposals and detailed
investigations how top quarks, and specifically associated polarization
phenomena,
can serve as probes  of (non) Standard Model interactions \cite{Zerwas} -
\cite{ArSe}.
In particular arguments have been given and proposals have been made to use
top quarks
produced in $e^+e^-$ collisions as probes of CP-violating interactions
(\cite{BNOS} - \cite{Wien}) beyond the
Kobayashi-Maskawa (KM) mechanism \cite{Ko}.

\noindent In this article we investigate, in continuation of
\cite{BO}, the effects on $\bar{t}t$
production of CP-violating interactions from an extended Higgs sector and
from the minimal
supersymmetric extension of the SM, propose observables for tracing these
effects and
calculate their expectation values. The new features of this article are:
we take into
account the possibility of longitudinally polarized electron beams which
enhance some of the
effects, and we propose and study optimized observables with maximized
sensitivity
to CP effects in the channels discussed below. (Some simple  correlations
 were calculated in \cite{Rind} for polarized $e^-$
beams in terms of CP-violating dipole form factors.)

\noindent The paper is organized as follows: In section 2 we discuss how
CP-violating effects
in $e^+e^-\to \bar{t}t$  manifest themselves in top spin-momentum
correlations and give
 observables to detect these correlations in "semihadronic" final states.
 The construction of the optimized observables is done in the appendix. In
sections 3 and 4
 we compute the expectation values of these observables and their
sensitivities within the
 above-mentioned models. Section 5 contains our conclusions.

\bigskip

\def\OObar#1{\vbox{\hbox{$\scriptscriptstyle
     (\vbox{\hrule width 5pt})$}\vskip-5pt \hbox{${\cal O}_#1$} }}

\font\tenbold=cmmib10
\newfam\boldfam \def\boldit{\fam\boldfam\tenbold}
\textfont\boldfam=\tenbold

\def\eplus{{\boldit e}_+}
\def\eminus{{\boldit e}_-}

\def\ehat{\hat {\boldit e}_+}

\def\kt{{\boldit k}_{\rm t}}
\def\ktbar{{\boldit k}_{\rm \bar t}}

\def\khat{\hat{\boldit k} }

\def\t{{\rm t}}
\def\tbar{{\rm \bar t}}

\def\qXbar{\hat{\boldit q}_{\bar X}}
\def\qX{\hat{\boldit q}_{X}}

\def\qLplus{{\hat{\boldit q}_+}}
\def\qLminus{{\hat{\boldit q}_-}}

\def\qplus{{\hat{\boldit q}}^{\ast}_+}
\def\qminus{{\hat{\boldit q}}^{\ast}_-}

\def\qplusminus{{\hat{\boldit q}}^{\ast}_\pm}

\section{Observables}

In the following  we consider the production of a top quark pair via the
collision of
an unpolarized positron  beam and a longitudinally polarized electron beam:

\begin{equation}
 {\rm e}^+ (\eplus)  +  {\rm e}^- (\eminus, p)  \quad \to \quad
\t( \kt) +  \tbar (\ktbar).
\label{TheReaction}
\end{equation}

\noindent Here $p$ is the longitudinal polarization of the electron beam
($p=1$ refers to right handed electrons).
We are interested in reactions with semileptonic t decay and non-leptonic
$\tbar$
decay and vice versa:

\begin{equation}
\t  \enskip  \tbar  \quad \to \quad \ell^+({\boldit q}_+) + \nu_\ell + {\rm b}
   + \overline{ X}_{\rm had}({\boldit q}_{\bar X}),
\label{TopDecay}
\end{equation}

\begin{equation}
  \t  \enskip  \tbar \quad \to \quad  X_{\rm had}({\boldit q}_X) +
\ell^-({\boldit
q}_-) +  \bar \nu_\ell     + {\rm \bar b} \  \ ,
\label{AntiTopDecay}
\end{equation}

\noindent  where the 3-momenta in eqs. (1) - (3)
refer to the $e^+ e^-$ c. m. frame. \\
\noindent If non-standard CP-violating interactions exist
they can affect the $\bar{t}t$ production
and decay vertices. Quantum mechanical interference of the CP-even and -odd
parts
of the amplitudes for the above reactions then lead to the correlations
which we are
after. For the models of sect.3,4 below it has been shown {\cite{BO}} that they
induce a CP-violating form factor  in $t$ (and $\bar{t}$) decay which
is  smaller than the electric and weak dipole form
factors of the top generated in the  production
vertex. Therefore we consider below observables which are predominantly
sensitive to these form factors. \\
\noindent The SM extensions of sect. 3,4 lead to electric and weak dipole
form factors
$d^{\gamma,Z}_t(s)$ which can have imaginary (i.e., absorptive) parts. The
real parts
${\rm Re} d_t^{\gamma,Z}$  induce a difference in the $t$ and $\bar{t}$
polarizations
orthogonal to the scattering plane of reaction (1). Non-zero absorptive parts
${\rm Im} d_t^{\gamma,Z}$ lead to a difference in the $t$ and $\bar{t}$
polarizations
along the top direction of flight. \\
The class of events (2), (3) is highly suited
to trace these spin-momentum correlations in the
$\bar{t}t$ production vertex through final state momentum correlations:
{}From the hadronic momentum in (\ref{TopDecay}),(\ref{AntiTopDecay}) one can
reconstruct the $\tbar$ and $t$ momentum, respectively and hence the rest
frames
of these quarks. Moreover the scattering plane of the reaction
(\ref{TheReaction})
can be determined in each event.
The extremely short life time of the top quark implies that the top
polarization is essentially undisturbed by hadronization effects and
can be analyzed by its parity-violating weak decay $t \to b + W.$ Further,
the charged lepton from semileptonic top decay is known to be by far the best
analyzer of the top spin {\cite{CJK}. \\
\noindent Therefore
the observables which will be discussed below are chosen to be
functions of the directions of the
hadronic system from top decay, of the charged lepton momentum, of
the positron beam direction, and of the c.m. energy $\sqrt s$.
In a previous article \cite{BO}
we have used observables being functions of the  lepton momenta in the
laboratory
system.  Aiming at the optimization of those results we are  lead to
observables involving the lepton unit momenta $\qplusminus$ in the
corresponding top
rest frames, which are directly accessible in the processes considered here.\\
\noindent The CP asymmetries which we discuss
are differences of expectation values

\begin{equation}
{\cal A} \quad = \quad \langle{\cal O}_+ (s,\qplus, \qXbar, \ehat)\rangle -
              \langle{\cal O}_- (s,\qminus, \qX, \ehat)\rangle \ \ ,
\label{GeneralForm}
\end{equation}

\noindent where the mean values refer to events
(\ref{TopDecay}), (\ref{AntiTopDecay}) respectively. The observable ${\cal
O}_-$ is
defined to be the CP image of ${\cal O}_+$; that is, it is obtained from
${\cal O}_+$
by the substitutions $\qXbar \to -\qX$, $\qplus \to -\qminus$, $\ehat \to
\ehat$. \\
In the sections below we shall plot the ratios

\begin{equation}
 r \quad = \quad {\langle{\cal O}_+\rangle - \langle{\cal O}_- \rangle \over
  \Delta{\cal O}_+}
 \label{Sensitivity}
\end{equation}

\noindent where $\Delta{\cal O} = \sqrt{\langle{\cal O}^2\rangle
-\langle{\cal O}\rangle^2}$
is the width of the distribution of ${\cal O}$. For the observables used
in this paper we have  $\Delta{\cal O} \approx \sqrt{\langle{\cal O}^2\rangle}$
and $\Delta{\cal O}_+\simeq \Delta{\cal O}_-$. The absolute value $|r|$ is
a measure of
the sensitivity of a correlation. The corresponding signal-to-noise ratio
is given by $S_{\cal A } = |r| \sqrt N / \sqrt 2$, where $N$ is the number
of events
of type (2) or (3).\\
\noindent In the appendix we derive explicit expressions for observables of
the above
type which lead to the largest possible sensitivity $|r|$. In addition to the
functional dependence exhibited in (\ref{GeneralForm}) they depend also on
the degree of beam polarization. At first sight these optimized observables
may appear
rather unhandy. Therefore we derive from the formulae  given in the
appendix  two
observables -- which are sensitive to dispersive
and absorptive CP effects, respectively --- with a simpler structure:

\begin{equation}
{\cal O}_+^{\rm Re} \quad = \quad  (\qXbar \times \qplus)\cdot\ehat \  \ ,
\label{ObsRe}
\end{equation}

\begin{equation}
{\cal O}_+^{\rm Im} \quad = \quad
- [1 + ({\scriptstyle\sqrt s \over \scriptstyle 2m_t} - 1)
(\qXbar\cdot\ehat)^2]\qplus\cdot\qXbar
 + {\scriptstyle\sqrt s \over\scriptstyle 2m_t} \enskip\qXbar\cdot\ehat
 \enskip \qplus\cdot\ehat  \ \ ,
\label{ObsIm}
\end{equation}

\noindent where $m_t$ denotes the top mass.
The observables ${\cal O}_-$ are obtained from  the corresponding
${\cal O}_+$ by the substitutions given below eq. (\ref{GeneralForm}). Eqs.
(\ref{ObsRe}),(\ref{ObsIm}) do not depend on the beam polarization degree
$p$. \\
\noindent In the narrow width approximation for
$\bar{t}t$ production and decay it can be shown (see {\cite{BNOS, BO}}) that
the
resulting asymmetries ${\cal A}^{\rm Re},{\cal A}^{\rm Im}$ are
predominantly sensitive
to CP violation in the $\bar{t}t$ production amplitude: ${\cal A}^{\rm Re}$
traces non-zero
 ${\rm Re} d_t^{\gamma,Z}$ whereas  ${\cal A}^{\rm Im}$ receives
contributions from the
 absorptive parts of these form factors.\\
 \noindent From the  structure of the observables
(\ref{ObsRe}),(\ref{ObsIm}) we can read off
whether the corresponding asymmetries depend on the polarization of the
electron beam:
Since ${\cal O}_\pm^{\rm Re}$ is linear in the beam direction $\ehat$ its
expectation value
is proportional to the vector polarization of the intermediate virtual
$\gamma/Z$ boson which
depends on $p$.  We find that the
corresponding ratio $r$ is enhanced by a factor of
about  two if the electron beam is  fully polarized as compared to the case
$ p=0$ (see
below). On the other hand
${\cal O}_\pm^{\rm Im}$  is bilinear in $\ehat$, leading to an asymmetry
which is blind
to the beam polarization.\\
\noindent In the following we shall always consider phase space cuts which
are CP-symmetric.
When the ${\rm e}^+ {\rm e}^-$ beams are unpolarized (or transversely
polarized) the
asymmetries (\ref{GeneralForm}) can be classified as being odd under a CP
transformation.
This means that contributions to $\langle {\cal O}_\pm \rangle$ from
CP--invariant interactions  cancel in the difference.
If the electron beam is longitudinally polarized the initial $e^+e^-$ state
is no longer
CP-symmetric in its c.m. frame and the CP classification no longer applies.
Contributions
from CP-conserving interactions can, in principle, contaminate ${\cal A}$
if $p\neq 0$.
However, in practice, this is not a problem for the following reasons: i)
At a high luminosity linear collider CP-violating
or -conserving interactions must induce ratios $|r|> 0.01$
in order to have a chance to be detectable. Effects in ${\cal A}$ at the
per mill level are swamped by the statistical fluctuations (and probably
overwhelmed by
systematic errors in an experiment). ii) To leading order in the
electroweak couplings
the contamination problem does not arise for the reaction (1) since the
intermediate
virtual photon or Z boson is still a CP eigenstate in the c.m. frame for
arbitrary
polarization $p$. iii)  The asymmetry ${\cal A}^{\rm Re}$ and the
corresponding optimized
version of the appendix are T-odd, that is, odd under reversal of momenta.
Hence only
absorptive parts of the CP-conserving component of the amplitude, which
arise at 1-loop order, can contribute. As to SM interactions only
absorptive parts of
electroweak box contributions to (1) are relevant in view of ii). They can
be estimated to be
of  the order of $\pi
\times e^2/(16{\pi}^2) \simeq 0.2 \%$ and are therefore negligible. (It
might nevertheless be
interesting to investigate these contributions in detail.) iv) The
asymmetry ${\cal A}^{\rm
Im}$ and the corresponding  optimized version of the appendix are T-even.
For $p\neq 0$ the
most important  CP-conserving contribution to these quantities comes from
helicity flip
bremsstrahlung off the initial $e^-$ or e$^+$. This effect has been
calculated in
{\cite{Brems}} and was shown to be small. For our asymmetries it leads to a
contamination of a
few per mill at most.

\bigskip

\section{Two--Higgs--doublet extensions \newline
    of the Standard Model  }

Only new CP-violating interactions beyond the Kobayashi Maskawa phase can
induce
recognizable effects in asymmetries of the form (\ref{GeneralForm}).
An intriguing interaction of this type is  provided already by
two--Higgs--doublet extensions of the SM with explicit CP violation in the
scalar potential. These models contain in their spectrum three
physical neutral Higgs
bosons  $\varphi_j (j=1,2,3)$ which no longer have a definite CP parity.
In these models the
Yukawa couplings of the $\varphi_j$ to the top quark are given by {\cite{BSP}}:

\begin{equation}
{\cal L}_Y  = - \left(\sqrt 2 G_F\right)^{1\over2}
\sum^3_{j=1} \left[
a_{j{\rm t}} m_{\rm t} {\rm \bar t}{\rm t}  +
\tilde a_{j {\rm t}} m_t {\rm \bar t}
i\gamma_5
{\rm t}
\right] \varphi_j
\label{Yuk}
\end{equation}

\noindent
where $m_t$ is the mass of the top quark,

\begin{equation}
a_{j{\rm t}} = d_{2j} / \sin \beta, \quad
  \tilde a_{j{\rm t}} = - d_{3j}  \cot \beta,
 \end{equation}

\noindent   $\tan \beta = {v_2/ v_1}$ is the  ratio of vacuum
expectation values of the two doublets, and $d_{2j}, d_{3j}$
are the matrix elements of a $3\times 3$ orthogonal matrix. \\
\noindent The Yukawa couplings (\ref{Yuk}) generate  electric and weak
dipole form
factors of the top quark at one loop order which have absorptive parts for
$s \ge 4 m_t^2$.
We shall assume that at least one of the $\varphi$'s is light; for definiteness
we take
$m_{\varphi_1} <\hskip-3pt< m_{\varphi_{2,3}}$.  Heavy Higgs bosons lead to
very small
effects in the   asymmetries of sect. 1 -- contrary to the case of resonant
$\varphi_j$
production  \cite{BeBra}.)
In the parameter range which we
shall explore below the asymmetries (\ref{GeneralForm}) are, to a good
approximation,  proportional to

\begin{equation}
\gamma_{\rm CP} \quad \equiv\quad  d_{21} d_{31} \cot \beta/\sin\beta.
\end{equation}

\noindent One may define CP violation in the neutral Higgs sector to be
maximal if
$|d_{i1}| = 1/\sqrt 3$ for i=1,2,3.
K and B meson data indicate that
$\tan \beta
\vbox{\hbox{$\scriptstyle>$} \vskip-11pt\hbox{$\scriptstyle\sim$}}
0.3$. This translates  into the rather loose upper bound:
$|\gamma_{\rm CP}|
\vbox{\hbox{$\scriptstyle<$}\vskip-11pt\hbox{$\scriptstyle\sim$}} 4.$
This bound is not in conflict with the present
experimental upper bounds on the electric dipole moment of the neutron
\cite{Neutron}
and of the electron {\cite{Commins}}.
Here we are interested in the case where the pseudoscalar coupling
$\tilde a_{1{\rm t}}$ is not severely suppressed.
If $\tan \beta >\hskip-3pt>1$ then   the effects, which we study in this paper,
become too small for being observable.
In the following we choose $\gamma_{\rm CP} = 2$, $m_{\rm t} =
180 \hskip 3pt{\rm GeV},$  and $m_{\varphi_1} = 100 \hskip 3pt{\rm GeV}.$

\noindent For these parameters  the   ratios $r$
which correspond to the observables (\ref{ObsRe}) and (\ref{ObsIm}),
respectively,  are plotted  in Figs. 1a,b as functions of the c.m. energy
using the
dipole form factors as obtained in {\cite{BSP}}.
{}From the figures one can infer the statistical sensitivities in
straightforward
fashion.

\noindent The asymmetry  ${\cal A}^{\rm Re}$ is proportional
to the real part of the dipole form factors. The corresponding ratio $r$ has
its
maximum near the $t\bar{t}$ threshold, becomes zero around
$\sqrt s  \approx  500 \hskip 3pt{\rm
GeV}$ (the zero originates from the real part of the form factors),
and increases again. Beyond 1.2 TeV it eventually decreases.
We find, as expected,  a strong dependence of $r$ on the
longitudinal polarization of the electron beam: For $p=\pm 1$ the sensitivity
is more than twice
as large as in the unpolarized case.

\noindent The difference  ${\cal A}^{\rm Im}$
projects onto the imaginary parts of the form factors. It leads to a ratio $r$
which reaches its maximum  of 2 percent  at about 500 GeV and decreases
moderately at
higher energies.
As mentioned above, the beam polarization has no effect on this asymmetry.

\noindent Instead of (\ref{ObsRe}) and (\ref{ObsIm}) we may use observables
with optimized signal-to-noise ratio.
This is done in appendix A.  We find that the corresponding sensitivities
increase by about 30 percent as compared to those given by Figs. 1a,b.
Fig. 3b shows that the absorptive asymmetry (\ref{ratio}) has the highest
sensitivity to $\gamma_{CP}$ and that it depends only weakly on the beam
polarization.
The maximal sensitivity is reached at $\sqrt s \simeq$ 450 GeV where
$|r_2|$ = 2.4$\%$ (2$\%$) if $p$=$\pm$1 (0). In order to detect this as a 3
s.d. effect
one would need 31000 (45000) events of the type (\ref{TopDecay}) and of
(\ref{AntiTopDecay}).
Since (\ref{TopDecay}) and  (\ref{AntiTopDecay}) correspond each to 2/9 of the
$\bar{t}t$ events this would require about 4 years of data collection with
a linear collider
having a luminosity of ${\cal L}= 5\times 10^{33}/cm^2s$. If the
CP-violating effect is
larger,
say $\gamma_{\rm CP}$=4, then only 1/4 of these events would be needed for
a 3 s.d. effect.

\noindent  In general the asymmetries become smaller with increasing Higgs
mass.
If we keep the CP-violating coupling $\gamma_{\rm CP}$ = 2 but change the Higgs
mass to $m_{\varphi}$  =
200 GeV  then the value of $|r|$ and   $|r_1|$ of Fig.1a and Fig.3a,
respectively,
at $\sqrt s \approx$ 384 GeV (which is the location of the
maximal value close to threshold)  and the maximal values of $|r|$ of
Fig.1b and $|r_2|$ of Fig.3b at $\sqrt s \approx$ 450 GeV
are reduced by a factor of about 0.7.

\noindent In summary, for light Higgs masses $m_{\varphi}<$200 GeV and sizable
CP-violating coupling $\gamma_{\rm CP}>$2 there is a chance to see Higgs
sector CP violation
as a virtual effect in $\bar{t}t$ production. A light Higgs
particle $\varphi$ would also
be produced at a linear collider. A consequence of $\varphi$ not being CP
eigenstate
would be a CP violation effect in the $\varphi$ fermion-antifermion
amplitude at Born level which could be detected in $\varphi\to\tau^+\tau^-$
\cite{BBra}.
(For further checks of the CP properties of Higgs particles, see \cite{DK,
Kremer}.)

\bigskip

\section{Minimal supersymmetric extension of the Standard Model}
It is well known that in the
minimal supersymmetric extension of the SM with two Higgs doublets
CP-violating terms are
absent in the Higgs potential. In the supersymmetric case neutral Higgs
sector CP violation
requires at least one additional scalar field, for instance a singlet.
However, already in the
minimal
 supersymmetric extension of the SM  additional  CP--violating
phases (besides the KM phase) can be present in the Majorana mass terms,
e.g. of the
gluinos, and in the squark (and slepton) mass matrices.
For mass eigenstates  these phases
then appear in the ${\rm t}\tilde{\rm t}-\rm gluino$  couplings in the form
(flavour mixing is ignored)

 \begin{equation}
 {\cal L}_{\tilde {\rm t} {\rm t} \lambda} =
 i\sqrt 2 \enskip g_{\rm\scriptscriptstyle QCD} \left\{
 e^{i\phi_{\rm t}} \enskip\tilde{\rm t} _L^*  T^a ( \bar\lambda^a {\rm t}_L)
 + e^{-i\phi_{\rm t}} \enskip\tilde {\rm t}^*_R T^a (\bar \lambda^a {\rm
t}_R) \right\} +
h.c.
\label{LSusy}
\end{equation}

 \noindent with $\phi_{\rm t} = \phi_\lambda - \phi_{\tilde {\rm t}}$ and

 \begin{eqnarray}
 \tilde {\rm t}_L \quad &= \quad \phantom{-}\tilde {\rm t}_1
\cos\alpha_{\rm t} +
 \tilde {\rm t}_2
 \sin \alpha_{\rm t},\\
 \tilde {\rm t}_R \quad &= \quad -\tilde {\rm t}_1 \sin \alpha_{\rm t} +
\tilde {\rm t}_2
 \cos\alpha_{\rm t},
 \end{eqnarray}

 \noindent where $\tilde {\rm t}_{1,2}$ denote the fields corresponding to mass
eigenstates.
 If $\phi_{\rm t}$ is ``flavour-universal", i.e. is the same for the
 $\tilde u, \tilde c, \tilde t$
 squarks, then the experimental upper bound on the electric dipole moment
of the neutron  puts a constraint on
the magnitude of $\phi_{\rm t}$ ( see, e.g. the reviews \cite{GRIM,HE}),
depending on the
magnitude of the gluino and squark masses. If $m_{\lambda}, m_{\tilde u},
m_{\tilde d}$ are
close to their present experimental lower bounds of about 200 GeV
{\cite{CDF}} (this bound
holds only if $m_{\lambda} = m_{\tilde q}$)   then
$|\sin(\phi_{\rm t})|
\leq 0.1.$ However, for masses  larger than 500 GeV this constraint disappears.
In any case  $\phi_{\tilde{\rm t}}$ may a priori  be much larger in
magnitude  than
$\phi_{\tilde u}.$
In order to investigate the
maximally possible magnitude of the effects  we shall put
$\sin(2\phi_{\rm t}) = 1.$ \\
\noindent In order that the interaction (\ref{LSusy}) induces CP--violating
dipole form factors in the
${\rm t}{\rm \bar t}$ production amplitude
$\tilde{\rm t}_{1,2}$ must not be mass--degenerate. We use the form factor
formulae
 of (\cite{BO}). In order to exhibit the typical size of the effects we
assume maximal
mixing, take the mass of
$\tilde{\rm t}_2$ to be $m_2 = 400 \hskip 3pt{\rm GeV}$ and those of
$\tilde{\rm t}_1$
in the vicinity of $m_t$, to wit: $m_1 = 190$ GeV. The gluino mass is chosen to
be $m_{\lambda} = 150$ GeV which is in accord with \cite{CDF}.
For longitudinal electron
polarization $p=0,\pm 1$
the  ratios $r$ corresponding to the observables (\ref{ObsRe}),
(\ref{ObsIm}) are
plotted in Figs. 2a,b. As expected, the sensitivity of
${\cal A}^{\rm Re}$
is large around the threshold
for  $\tilde{\rm t}_1\tilde{\rm t}_1^*$ production.
(In the close vicinity of a squark threshold
our results may become unreliable due to resonance
effects).
Above a zero  at about 750 GeV there is another local maximum
of  $|r| \approx= 0.6\%$ at
$\sqrt s \approx$ 1.2 TeV for $p$= -1, which is slightly larger
than the maximum near threshold.
The absorptive asymmetry ${\cal A}^{\rm Im}$  is largest in
the region where the dispersive asymmetry is very small. For the parameters
above the
maximal sensitivity is 0.5 $\%$ at $\sqrt s \approx$ 600 GeV
independent of the polarization. \\
\noindent We have evaluated the ratios $r$ also for other mixing angles,
squark and
gluino masses, requiring that $m_{\lambda}, m_{1,2}  \ge$ 150 GeV. We have
not found
values of $|r|$ larger than the ones given above. (The location of the
maxima do of course
change.)  If one uses the optimal observables (\ref{Opt}) given in the
appendix the
corresponding sensitivities increase by about 30$\%$ as compared to those
displayed in
Figs. 2a,b. That is, even with these quantities we get ratios $|r|
\vbox{\hbox{$\scriptstyle<$}\vskip-11pt\hbox{$\scriptstyle\sim$}}$0.01.

\noindent A number of supersymmetry induced  CP correlations and asymmetries
for $\bar{t}t$ production and decay were also studied in refs.
{\cite{GRA,Christ,Wien}}. The
observables used in these references have a smaller sensitivity to
supersymmetric CP violation
than the ones which we have proposed and evaluated here.

\section{Conclusions}
\noindent We have constructed optimized observables to search for CP
violation in
$\bar{t}t$ production using "semihadronic" final states. We have taken into
account the
possibility of both unpolarized and longitudinally polarized electron
beams, and have
calculated the expectation values of these observables for Higgs boson-induced
and supersymmetry-induced CP violation. We have found that the
optimal correlation for tracing dispersive CP effects is enhanced with
longitudinally
polarized electron beams, whereas the best correlation for probing
absorptive CP effects
is almost independent of the polarization degree. Moreover we have shown
that the
the latter correlation has the highest sensitivity to CP-violating Higgs
boson exchange.

\noindent CP phases in gluino exchange lead to ratios
$|r|\vbox{\hbox{$\scriptstyle<$}\vskip-11pt\hbox{$\scriptstyle\sim$}}$1$\%$.
These
effects are too small to be detectable as a, say 3  s.d. effect at a linear
collider
with integrated luminosity of 50 ${\rm (fb)^{-1}}$/year.
On the other hand if a light Higgs boson with mass
$m_{\varphi}<$200 GeV and sizeable CP-violating couplings to top quarks
exists then,
as shown above, an effect could be seen with the absorptive observable
proposed in
this paper.

\begin{appendix}

\section {Optimal observables}

In this appendix we give explicit formulae for the optimized  observables
which trace effects of CP--violating dipole moments in the reactions (1),
(2), (3).
The method of optimizing the
signal--to--noise  ratio of observables has  proven to be a powerful tool
for CP studies in tau-pair production at LEP {\cite{Over, Opal}}. In order
to illustrate the
procedure  we consider a differential cross section which is  of the
schematic form ${\rm d}
\sigma = {\rm d}
\sigma_0 + \lambda {\rm d} \sigma_1$. Phase space variables which
are not measured in a particular experiment are understood to be  integrated
out. The parameter $\lambda$ is
assumed to be small so that possible higher order terms in $\lambda$ can be
neglected. If one
wants to measure
$\lambda$  by the mean value of an appropriate observable one can show
{\cite{AS}}
(see also {\cite{Davier}})
that the optimal observable, that is, the one with the minimal statistical
error is given by
\begin{equation}
{\cal O} \quad = \quad { {\rm d} \sigma_1 /
                        {\rm d} \sigma_0 } .
\label{OptObs}
\end{equation}

\noindent This observable remains optimal in the presence of phase space cuts.
 A generalization to  the case of several parameters
was given in {\cite{Diehl}}. In our case  ${\rm d} \sigma_0$
denotes the CP--conserving part of the differential cross section of the
above reactions, while
$\lambda {\rm d} \sigma_1$ is the CP--violating contribution.

\noindent In our case the
CP-violating term of the the differential cross section is linear in
the real and the imaginary parts of the electric and  the weak dipole form
factor of the top
quark; i.e., it is of the form
$\sum \lambda_i {\rm d} \sigma_1^i$. Each of these four couplings, which
are a priori
unknown, can be
measured with  observables ${\cal O}(i)  =  { {\rm d} \sigma_1^i /{\rm d}
\sigma_0}$ .

\noindent For practical purposes it is sufficient to use the tree level
approximation of
the  transition matrix element squared  when writing down the observables
${\cal O}(i)$. We
give expressions for the final states (2) and (3)
under the proviso that, for given c.m. energy (and top mass),
the ${\cal O}(i)$ depend
only on the the momentum direction $\qplus$ of the charged lepton in the
rest system of the
top quark and on the $\bar{t}$ momentum direction in the laboratory
frame when considering the channels (2), and
analogously for the final states (3).  Then the analytic form of the
optimized observables is not too complicated; partly because in semileptonic
top
decays the energy and the direction of flight of the lepton are not
correlated to lowest order,
and because the
terms which involve correlations between the lepton
momenta
$\qplus$ and $\qminus$ originating from $t \bar{t}$
spin correlations  do not contribute.
The optimized observables have the following form (as before the labels
$\pm$ refer to
the final states (2) and (3), respectively):

\begin{equation}
{ \cal O}_\pm(i) =
{ \displaystyle
\sum_{B_1,B_2 = \gamma,Z}
{ {R_\pm^1}(B_1,B_2)_i \over (s-M_{B_1}^2  ) (s-M_{B_2}^2  ) }
\over \displaystyle
\sum_{B_1,B_2 = \gamma,Z}
{ {R_\pm^0}(B_1,B_2) \over (s-M_{B_1}^2  ) (s-M_{B_2}^2  ) }
} \       \ .
\label{Opt}
\end{equation}

\noindent The $R_\pm^0$ are proportional to the respective Standard Model
Born matrix elements. (Overall multiplicative constants or s-dependent
terms are irrelevant and
can be neglected.) For $R_+^0$ we have:

\begin{eqnarray}
\lefteqn
 { {R_+^0}(B_1,B_2) \quad =} \nonumber \\
&&\left( v^{B_1}_e (v^{B_2}_e -p a^{B_2}_e) + a^{B_1}_e (a^{B_2}_e - p
v^{B_2}_e) \right)
     \nonumber \\
&& \quad\quad \cdot \Bigl [  v^{B_1}_t v^{B_2}_t ( k_0^2 + m_t^2 + k^2
(\khat\cdot\ehat)^2 )
			           + a^{B_1}_t a^{B_2}_t k^2 ( 1+(\khat\cdot\ehat)^2) \nonumber \\
&&  \quad\quad   +  (v^{B_1}_t a^{B_2}_t + a^{B_1}_t v^{B_2}_t )
              k ( \khat\cdot\qplus ( k_0 + (k_0{\scriptstyle - }m_t)
(\khat\cdot\ehat)^2 )
              + m_t \khat\cdot\ehat \enskip \ehat\cdot\qplus ) \Bigr ]
\nonumber \\
&&+\left( v^{B_1}_e (a^{B_2}_e -p v^{B_2}_e) + a^{B_1}_e (v^{B_2}_e - p
a^{B_2}_e) \right)\\
&&\quad\quad \cdot\Bigl[ v^{B_1}_t v^{B_2}_t 2k_0 ( m_t \ehat\cdot\qplus +
(k_0{\scriptstyle - } m_t)\khat\cdot\ehat \khat\cdot\qplus )
\nonumber \\
&& \quad\quad+ a^{B_1}_t a^{B_2}_t 2 k^2 \khat\cdot\ehat \enskip
\khat\cdot\qplus
\quad
 + (v^{B_1}_t a^{B_2}_t + a^{B_1}_t v^{B_2}_t) 2 k_0 k \khat\cdot\ehat
\Bigr ] \nonumber
\label{Standard}
 \end{eqnarray}

\noindent Here $\khat$ denotes the direction of the antitop quark (to be
identified with $\qXbar$) in the laboratory frame, $ k = \sqrt{s/4 -
m_t^2}$, and
$k_0={\sqrt s}/2$. The neutral electroweak
couplings of $f= e,t$ are given by

\begin{eqnarray}
&v_f^\gamma = Q_f \cdot e \hskip 2cm
&v_f^Z = {T_f^3 -2 Q_f \sin^2 \theta_W \over 2 \sin \theta_W \cos \theta_W
} \cdot e \\
&a_f^\gamma = 0  \hskip 2cm
&a_f^Z = {T_f^3  \over 2 \sin \theta_W \cos \theta_W } \cdot e \nonumber
 \label{NC}
\end{eqnarray}

\noindent The expression for $R_-^0$ is obtained from eq.(16) by the
substitution
$\qplus \to -\qminus$, and in experimental applications  one should also
replace the
$\bar{t}$ unit momentum $\khat$ by $-\khat_{top} = - \qX$.

\noindent The CP-violating part of the respective differential cross
section depends on the real
and imaginary parts of the electric and weak dipole form factors.
CP-violating form factors in the $t$ and $\bar{t}$ decay vertices play no
role in our case, as
shown in {\cite{BNOS,BO}}. For the final state (2) the
CP-violating matrix element squared is proportional
to

\begin{eqnarray}
\lefteqn
 { {R_+^1}(B_1,B_2) \quad =} \nonumber \\
& &\phantom{+}\left( v^{B_1}_e (a^{B_2}_e -p v^{B_2}_e) + a^{B_1}_e
(v^{B_2}_e -p a^{B_2}_e)
\right)
   \left(  v^{B_1}_t {\rm Re} d_t^{B_2} +v^{B_2}_t {\rm Re} d_t^{B_1} \right)
     \nonumber \\
& & \quad\quad \cdot  k_0^2 k ( \khat \times \qplus )\cdot\ehat
      \nonumber \\
& &+\left( v^{B_1}_e (v^{B_2}_e -p a^{B_2}_e) + a^{B_1}_e (a^{B_2}_e -p
v^{B_2}_e) \right )
   \left(  a^{B_1}_t {\rm Re} d_t^{B_2} + a^{B_2}_t {\rm Re} d_t^{B_1} \right)
     \nonumber \\
& & \quad\quad \cdot k_0 k^2 \enskip \khat\cdot\ehat \enskip
(\khat\times\qplus)\cdot\ehat
     \nonumber \\
& &+\left( v^{B_1}_e (v^{B_2}_e -p a^{B_2}_e) + a^{B_1}_e (a^{B_2}_e -p
v^{B_2}_e) \right )
   \left(  v^{B_1}_t {\rm Im} d_t^{B_2} + v^{B_2}_t {\rm Im} d_t^{B_1} \right)
     \nonumber \\
& & \quad\quad \cdot k_0 k\left [ - (m_t + (k_0{\scriptstyle - }m_t)
             (\khat\cdot\ehat)^2)\qplus\cdot\khat
          + k_0 \enskip\khat\cdot\ehat \enskip
				\qplus\cdot\ehat   \right]
    \nonumber \\
& &+\left( v^{B_1}_e (a^{B_2}_e -p v^{B_2}_e) + a^{B_1}_e (v^{B_2}_e -p
a^{B_2}_e) \right)
   \left(  a^{B_1}_t {\rm Im} d_t^{B_2} + a^{B_2}_t {\rm Im} d_t^{B_1} \right)
     \nonumber \\
& & \quad\quad \cdot k_0 k^2 \left[ \qplus\cdot\ehat
    - \khat\cdot\ehat \enskip \khat\cdot\qplus \right] \    \ .
 \label{CPodd}
\end{eqnarray}

\noindent The expression for $R_-^1$ is obtained from (\ref{CPodd}) by the
substitutions
given below eq.(17). The coefficients of ${\rm Re} d_t^{\gamma, Z}$ and
${\rm Im} d_t^{\gamma, Z}$  in  $R_\pm^1$ then define the ${R_\pm^1}_i$
which appear in
(\ref{Opt}). Note that  the
${\cal O}(i)$  are functions of the
polarization degree $p$ of the electron beam.

\noindent In the threshold domain the terms proportional to the vector
couplings $v_t$  are
the dominant  ones in $R_\pm^1$, while the axial couplings $a_t$ of the top
quark come with
an extra  p--wave suppression factor $k$. Keeping only the terms which are
dominant
-- as long as one is not too far above threshold -- one is led to the
observables
(\ref{ObsRe}),(\ref{ObsIm}) which do not depend on $p$. \\
\noindent In the following we concentrate on the Higgs models of sect.3 as
we have seen that
this type of CP-nonconservation can give larger effects than the SUSY phases
discussed in sect.4. For the models of sect.3 one has
$d_t^{\gamma}(s) \approx 3 d_t^Z(s)$ in the parameter range of interest.
Inserting this relation
into eq.(\ref{CPodd}) we can then write down two  optimized observables
which we denote by
${\cal O}_{\pm}(1)$ and ${\cal O}_{\pm}(2)$. They trace dispersive and
absorptive CP effects,
respectively. In Figs. 3a,b we have plotted the ratios

\begin{equation}
 r_i \quad  = \quad {\langle{\cal O}_+(i)\rangle - \langle{\cal O}_-(i)
\rangle \over
 \Delta{\cal O}_+(i) } \    \ (i = 1,2),
\label{ratio}
\end{equation}

\noindent as a function of the c.m. energy  for polarizations $ p = 0,\pm
1$. Comparing with
Figs. 1a,b one sees that at the energies where the respective sensitivities
are highest one
gains  in sensitivity by about 30 $\%$ when using the optimal observables
rather than those of
eqs. (\ref{ObsRe}), (\ref{ObsIm}). Moreover, since the dominant terms in
${\cal O}(1)$ are
linear in the  beam direction $\ehat$ whereas the dominant terms in ${\cal
O}(2)$
are bilinear in $\ehat$, it is clear that $r_1$ depends stronger on $p$
than $r_2$.

\noindent Figs. 3a,b show  that, when the $e^-$ beam is
maximally polarized, the sensitivity  of the optimized
observables  is the same for  $p=1$ and $p=-1$. This
result can be derived analytically for the ratios $r_{1,2}$ by
straightforward algebra.
The essential point is as follows:
Consider the quantity

\begin{equation}
C_p(B_1,B_2) \quad = \quad
{
v^{B_1}_e (a^{B_2}_e -p v^{B_2}_e) + a^{B_1}_e (v^{B_2}_e -p a^{B_2}_e)
\over
v^{B_1}_e (v^{B_2}_e -p a^{B_2}_e) + a^{B_1}_e (a^{B_2}_e -p v^{B_2}_e)
} \    \ ,
\label{Vec}
\end{equation}

\noindent where $B_1, B_2 = \gamma, {\rm Z}$. For $B_1 = B_2$
(\ref{Vec}) can be
interpreted as the vector polarization of the virtual $\gamma/ {\rm
Z}$--intermediate state.
It appears both in $R_\pm^0$ and $R_\pm^1$ when factoring out the term
which forms  the
denominator of eq.(\ref{Vec}).
If $p =\pm 1$ then
$C_p$ does not depend on coupling constants:

\begin{equation}
C_{p=\pm 1} (B_1,B_2) \quad = \quad \mp 1 \   \ .
\end{equation}

\noindent The ratios $r_{1,2}$ can be shown to be  even functions of
 $C_p(B_1,B_2)$ and therefore do not discriminate between $p=1$ and $p=-1$.
Of course if
 $|p| \neq$1 these ratios do depend on the sign of $p$.

\end{appendix}

\vfill\eject

{\Large\noindent Figure Captions}
\bigskip\bigskip\bigskip

{\noindent
Fig 1a: \quad Ratio $r$ for observable ${\cal O}^{\rm Re}_\pm$
evaluated with  parameters
$m_{\rm t} = 180$ GeV, $m_{\varphi_1} = 100$ GeV, and $\gamma_{\rm CP}$ = 2.
The longitudinal electron polarization is $p=0, \pm 1.$}

\bigskip
{\noindent
Fig 1b: \quad Ratio $r$  for observable ${\cal O}^{\rm Im}_\pm$
with parameters as in Fig. 1a.}

\bigskip

{\noindent
Fig 2a: \quad Ratio $r$ for observable ${\cal O}^{\rm Re}_\pm$
evaluated with parameters  $\sin 2\phi_{\rm t}=1,$
$m_{\rm t} = 180$ GeV, $m_{\lambda} = 150$ GeV, $m_1=190$ GeV, $m_2 = 400$
GeV, and
$\alpha_{\rm t} = \pi/4$  for electron polarization $p=0, \pm 1$. }
\bigskip

{\noindent
Fig 2b: \quad Ratio $r$ for observable ${\cal O}^{\rm Im}_\pm$
with  parameters   as in Fig 2a.}

\bigskip

{\noindent
Fig 3a: \quad Ratio $r_1$ defined in eq.(\ref{ratio}) evaluated  in the
Higgs model with
parameters as in Fig. 1a.}

\bigskip

{\noindent
Fig 3b: \quad Ratio $r_2$   defined in eq.(\ref{ratio}) evaluated
in the Higgs model with parameters as in Fig. 1a.}


\bigskip

\baselineskip=12pt

\begin{thebibliography}{99}
\bibitem{Zerwas}
"$e^+e^-$ Collisions at 500 GeV: The Physics Potential", ed. by
P.M. Zerwas, DESY orange reports DESY 92-123A,B and DESY 93-123C;\\
"Physics and Experiments with Linear Colliders", ed. by R. Orava,\\
P. Eerola, M. Nordberg, Singapore (1992)

\bibitem{KRZ}
 J.H. K\"uhn, A. Reiter, P. Zerwas: Nucl. Phys. B272 (1986) 560
\bibitem{Kroll}
  M. Anselmino, P. Kroll, B. Pire: Phys. Lett. B167 (1986) 113
\bibitem{Kuehn1}
M. Jezabek, J.H. K\"uhn: Phys. Lett. B329 (1994) 317
\bibitem{Kuehn2}
R. Harlander, M. Jezabek, J.H. K\"uhn, T. Teubner: Phys. Lett. B346 (1995) 137
\bibitem{Koerner}
J.G. K\"orner, A. Pilaftsis, M.M. Tung: Z. Phys. C63 (1994) 575;\\
S. Groote, J.G. K\"orner, M.M. Tung: Mainz preprint MZ-TH/95-09 (1995)
\bibitem{BNOS}
W. Bernreuther, O. Nachtmann, P. Overmann, T. Schr\"oder: \\
Nucl. Phys.   B388  (1992) 53; Erratum ibid. B406 (1993) 516
\bibitem{KLY}
G.L. Kane, G. A. Ladinsky, C.--P. Yuan: Phys. Rev. D45  (1991) 124
\bibitem{ArSe}
T. Arens, L.M. Sehgal: Nucl. Phys. B393 (1993) 46; Phys. Rev. D50 (1994) 4372
\bibitem{BSP}
W. Bernreuther,  T. Schr\"oder, T.N. Pham: Phys. Lett.  B279 (1992) 389
\bibitem{BO}
W. Bernreuther, P. Overmann: Z. Phys. C61 (1994) 599.
\bibitem{AS}
D. Atwood, A. Soni: Phys. Rev. D45 (1992) 2405
\bibitem{AS2}
D. Atwood, G. Eilam, A. Soni: Phys. Rev. Lett. 71 (1993) 492;\\
D. Atwood et al.: preprint SLAC-PUB-95-6981 (1995)
\bibitem{BM}
J.P. Ma, A. Brandenburg: Z. Phys. C56 (1992) 97; \\
A. Brandenburg, J.P. Ma: Phys. Lett. B298 (1993) 211
\bibitem{GRA}
B. Grzadkowski: Phys. Lett. B305 (1993) 384;\\
B. Grzadkowski, W.-Y. Keung: Phys. Lett. B316 (1993) 137
\bibitem{Rind}
F. Cuypers, S.D. Rindani: Phys. Lett. B343 (1995) 333;\\
S. Poulose, S.D. Rindani: Phys. Lett. B349 (1995) 379;
preprint hep-ph/9509299
\bibitem{Christ}
E. Christova, M. Fabbrichesi: Phys. Lett. B320 (1994) 299
\bibitem{Wien}
A. Bartl, E. Christova, W. Majerotto: Wien preprint HEPHY-PUB 624/95 (1995)
\bibitem{Ko}
M. Kobayashi, T. Maskawa:  Progr. Theor. Phys. 49  (1973) 652
\bibitem{Brems}
B. Ananthanarayan, S.D. Rindani: Phys. Rev. D52 (1995) 2684
\bibitem{CJK}
A. Czarnecki, M. Jezabek, J.H. K\"uhn: Nucl. Phys. B351 (1991) 70
\bibitem{BBra}
W. Bernreuther, A. Brandenburg: Phys. Lett. B314 (1993) 104
\bibitem{DK}
A. Djouadi, B. Kniehl, in: Ref. 1, DESY 93-123C (1993) p. 51
\bibitem{Kremer}
M. Kremer, J. K\"uhn, M. Stong, P. Zerwas: Z. Phys. C64 (1994) 21
\bibitem{GRIM}
 W. Grimus: Fortschr. Phys. 36 (1988) 201
 \bibitem{HE}
X.G. He, B.H.J. McKellar, S. Pakvasa: Int. J. Mod. Phys. A4 (1989) 5011
\bibitem{BeBra}
H. Anlauf, W. Bernreuther, A. Brandenburg: Phys. Rev. D52 (1995) 3803
\bibitem{Neutron}
K.F. Smith et al.: Phys. Lett. B234 (1990) 191
\bibitem{Commins}
E. D. Commins et al.: Phys. Rev. A50 (1994) 2960.
\bibitem{CDF}
S. Hagopian, in: "Proceedings of the XXVII Int. Conf. on High Energy Physics"
ed. by P.J. Bussey and I.G. Knowles, Bristol (1995), p. 809
\bibitem{Over}
P. Overmann: Dortmund preprint DO-TH 93-24 (1993)
\bibitem{Opal}
R. Akers et al. (OPAL collab.): Z. Phys. C66 (1995) 31
\bibitem{Davier}
M. Davier et al.: Phys. Lett B306 (1993) 411
\bibitem{Diehl}
M. Diehl and O. Nachtmann: Z. Phys. C62 (1994) 397
\end{thebibliography}
\end{document}